\documentclass[rmp,amsmath,amssymb, showkeys]{revtex4}
\usepackage{subfigure}
\usepackage{graphicx}
\usepackage{dcolumn}
\usepackage{bm}

\begin{document}


\title{Electron bunch injection at an angle into a laser wakefield}

\author{M.J.H. Luttikhof}
\email{m.j.h.luttikhof@tnw.utwente.nl}
\author{A.G. Khachatryan}
\author{F.A. van Goor}
\author{K.-J. Boller}
\affiliation{Faculty of Science and Technology and MESA+ Institute, University of Twente, P.O. Box 217, 7500 AE Enschede, The Netherlands}

\author{P. Mora}
\affiliation{Centre de Physique Th\'{e}orique, \'{E}cole Polytechnique, 91128 Palaiseau, France}

\date{\today}

\begin{abstract}
External injection of electron bunches longer than the plasma wavelength in a laser wakefield accelerator can lead to the generation of femtosecond ultrarelativistic bunches with a couple of percent energy spread. Extensive study has been done on external electron bunch (e.g. one generated by a photo-cathode rf linac) injection in a laser wakefield for different configurations. In this paper we investigate a new way of external injection where the electron bunch is injected at a small angle into the wakefield. This way one can avoid the ponderomotive scattering as well as the vacuum-plasma transition region, which tend to destroy the injected bunch. In our simulations, the effect of the laser pulse dynamics is also taken into account. It is shown that injection at an angle can provide compressed and accelerated electron bunches with less than 2\% energy spread. Another advantage of this scheme is that it has less stringent requirements in terms of the size of the injected bunch and there is the potential to trap more charge.
\end{abstract}


\maketitle

\newpage

\section{\label{sec:introduction}Introduction}
When an ultrashort laser pulse with high intensity propagates through a plasma, an electron density wave (laser wakefield) is generated due to the ponderomotive force exerted on the electrons \cite{tajima}. The electric fields formed in the plasma are huge. They have the potential to accelerate electrons to GeV in a plasma channel, of only a few centimeter. (see review article \cite{esarey}). In the all-optical regime, where electrons from the background plasma are trapped in the wakefield, acceleration of electrons has already been observed in experiments. An example is the ''bubble'' injection method proposed in \cite{pukhov} and demonstrated in 2004 \cite{geddes,faure,mangles}, where quasimonoenergetic electron bunches with an energy in the order of 100 MeV and an energy spread of a few percent were observed. More recently, by channeling the laser pulse through a 3.3 cm long plasma channel, electron bunches with an energy of an unprecedented 1 GeV and an energy spread of 2.5\% were generated \cite{leemans}. However the shot-to-shot reproducibility was still poor due to the strongly nonlinear dynamics of the laser-plasma interaction, which is rather sensitive to the shot-to-shot fluctuations in laser and plasma parameters. The reproducibility was considerably improved by employing a second, counterpropagating laser pulse \cite{faure2} to inject some electrons from the background plasma into the first accelerating region of the wakefield created by the first pulse. However a well-controlled laser wakefield accelerator remains still a big challenge.

More control might be obtained by using external electrons, instead of electrons from the background plasma, because the parameters of the injected bunch and the generated laser wakefield can be controlled independently. Also wakefields in the linear and weakly nonlinear regime can be used, providing a higher stability. It has been shown theoretically that the external injection of an electron bunch longer than the plasma wavelength in the laser wakefield can lead to the generation of femtosecond relativistic bunches with a relatively low energy spread \cite{khachatryan, khachatryan2, gordon, gorbunov, hubbard, lifschitz, esirkepov, kalmykov, khachatryan7, khachatryan6, khachatryan5, lifschitz2}. Such an electron bunch, which typically needs to have an energy of a few MeV, can be obtained from a standard radio-frequency (rf) photocathode linear accelerator. It is possible to compress this bunch to a length of about 100 $\mu$m and focus it down to a few tens of microns radius \cite{urbanus, irman}, which is comparable to the laser pulse spot radius in a channel-guided laser wakefield accelerator. The electron bunch can be injected in the wakefield formed in a plasma channel directly behind the laser pulse \cite{gorbunov, gordon, hubbard, lifschitz, kalmykov, khachatryan6, luttikhof}, but here two effects play a perturbing role.

The first effect is ponderomotive scattering of the injected bunch in vacuum \cite{khachatryan6}. To position the injected electron bunch behind the laser pulse at the entrance of the plasma channel, the bunch has to go through the pulse in the vacuum. The ponderomotive force of the laser pulse may scatter and eventually destroy the injected bunch in vacuum before it can be trapped and accelerated in the laser wakefield. The second effect is caused by the transition from vacuum to plasma \cite{luttikhof}. In this transition region, which typically has a length of the order of a couple of millimeters for a plasma channel, the plasma wavelength changes continuously, such that the injected electrons experience a rapidly changing wakefield, from focusing to defocusing. This way, a large fraction of the electrons can be scattered away in this transition region instead of reaching the regular wakefield.

There are two alternatives for external bunch injection into a laser wakefield, where these effects do not play a role. The first alternative is the injection of a low energy electron bunch in a plasma channel directly in front of the laser pulse \cite{khachatryan5, khachatryan, khachatryan2, esirkepov, khachatryan7}, such that the laser pulse will overtake the electron bunch in the plasma. In this case, when the bunch goes through the laser pulse, the excited laser wakefield roughly compensates the ponderomotive force, so that no ponderomotive scattering takes place in the plasma \cite{khachatryan,khachatryan2,khachatryan7}. Also the perturbing effect of the vacuum-plasma transition region can be avoided easily with a proper timing and, consequently, the injected bunch can be trapped, compressed and accelerated in the first accelerating bucket of the wakefield.

The second alternative, proposed in \cite{luttikhof} is the injection of an electron bunch at an angle. In this scheme, the bunch is injected at a small angle into a plasma channel directly behind the laser pulse, where a wakefield is generated. In this case both the ponderomotive scattering and the effect of the vacuum-plasma transition region are avoided.

In this article we present a detailed investigation of such injection at an angle into the laser wakefield. The novel injection method will be described in section \ref{sec:injection_at_an_angle}, and in section \ref{sec:laserdynamics} we will also look at the laser pulse dynamics in a plasma channel for typical parameters of the problem. In section \ref{sec:injectionatangledynamics} the effect of this dynamics on the trapping and acceleration of electrons injected at an angle is investigated. Section \ref{sec:summary} summarizes and discusses the results.

\section{\label{sec:injection_at_an_angle}Electron bunch injection at an angle}
\begin{figure}
\includegraphics[width=0.5\textwidth]{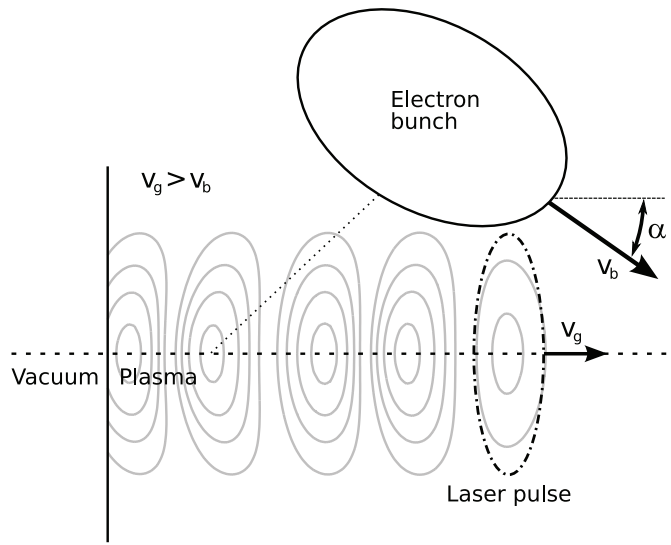}
\caption{\label{fig:schematic}Schematic drawing of injection at an angle into a laser wakefield. An electron bunch with velocity $v_b$ is injected into a laser wakefield with a small angle $\alpha$ with respect to the laser propagation axis.}
\end{figure}

In this section the working principal of injection of electron bunches at an angle is explained. A schematic drawing of the injection scheme is shown in figure \ref{fig:schematic}.
A low-energy electron bunch with typically a few MeV kinetic energy is generated by an rf photocathode linac and injected at a small angle, $\alpha$, directly behind a high-intensity laser pulse in a plasma with some offset relative to the propagation axis. This means that the electron bunch enters the wakefield from the transverse direction. It is preferable to inject the bunch into the wakefield close to the laser pulse because, here, the accelerating field is the largest and generates bunches with the highest quality, as was shown in \cite{khachatryan6}. The velocity of the electron bunch ($v_b$) is lower than the group velocity of the laser pulse ($v_g$), which means that the electron bunch slips backward in the frame moving with the pulse. The dotted line in figure \ref{fig:schematic} indicates the trajectory the bunch will follow in that frame. When the longitudinally accelerating and transversely focusing forces of the generated wakefield are sufficiently strong, and when also other parameters are suitably chosen, a large fraction of the injected bunch might be trapped, compressed and accelerated in the wakefield. In \cite{luttikhof} we proposed this method as a way to avoid ponderomotive scattering and the vacuum-plasma density transition as described above and we showed that the dynamics of injection at an angle look much like the dynamics of injection in front of the laser pulse \cite{khachatryan, khachatryan2,khachatryan7}. In both cases the injected electrons move gradually into the wakefield. In contrast, electrons injected on-axis behind the laser pulse suddenly experience a wakefield of which the strength depends on the injection phase \cite{gordon, khachatryan6}.

Besides avoiding ponderomotive scattering and the vacuum-plasma transition, which could also been done by injecting the electron bunch on-axis in front of the laser pulse, we show that there are additional advantages that make injection at an angle interesting. It is, for example, possible to use electron bunches of wider transverse size, which allows one to trap bunches with a higher charge in the wakefield. The longitudinal length of the electron bunch is not a critical issue, when a low energy spread is to be obtained. The injected bunch, from an rf linac, is typically a few times longer than the plasma wavelength $\lambda_p$ \cite{urbanus, irman}. Even longer injected bunches may provide low energy spread in the accelerated bunches with this injection scheme. This gives several compressed and accelerated bunches in the wakefield, separated by the distance $\lambda_p$. The trapping distance, an important quantity that determines the final energy spread of the bunch, can be made relatively small ($\sim$ 1 mm \cite{luttikhof}) by choosing larger angles. The wakefield should however be sufficiently strong to be able to also transversely trap electrons which are injected at larger angles.

\section{\label{sec:laserdynamics}Laser pulse dynamics}
In this section we will investigate the dynamics of the laser pulse, while it moves through a typical plasma channel. In our calculations we use a linearly polarized axially-symmetrical Gaussian laser pulse with the normalized amplitude, $a = e E_L/(m_e c \omega_L)$ \cite{esarey}, described as:
\begin{equation} \label{eq:laser}
a = a_0 \exp\left(-\frac{r^2}{w_0^2} - \frac{(z-v_g t)^2}{\sigma^2}\right),
\end{equation}
where $a_0$ is the normalized peak amplitude, $\sigma$ is the length of the pulse, $v_g$ is the group velocity of the pulse and $z$ and $r$ are the cylindrical coordinates. The laser pulse has initially a temporal full-width-at-half-maximum (FWHM) duration for the intensity of 50 fs (corresponding to $\sigma = 12.7$ $\mu$m) and is focused to a waist spotsize, $w_0$, of 30 $\mu$m. The Rayleigh length, $Z_R = \pi w_0^2/\lambda_L$, for such a pulse, with a wavelength ($\lambda_L$) of 800 nm, is 3.5 mm. A plasma channel is used to prevent diffraction of the pulse, so that it can be guided over several centimeters. We take a typical plasma density of 7 $\times$ 10$^{17}$ cm$^{-3}$, which corresponds to a plasma wavelength ($\lambda_p$) of 40 $\mu$m. Depending on the chosen peak amplitude, the peak power of the pulse is
\begin{equation}
 P[\textnormal{TW}] \approx 21.5 \times 10^{-3} (a_0 w_0/\lambda_L)^2 \approx 30.2 a_0^2.
\end{equation}
In comparison, the critical power for relativistic self-focusing is given by \cite{esarey3},
\begin{equation}
 P_c[\textnormal{TW}] \approx 17.4 \times 10^{-3} (\lambda_p/\lambda_L)^2 \approx 43.5.
\end{equation}
In this article we will restrict to weakly relativistic pulses with a normalized peak amplitude, $a_0$, of 0.6 and 0.8. The power corresponding to these pulses is, respectively, 10.9 TW and 19.3 TW, which is well below the critical power for self-focusing. However, the dynamics of the laser pulse can still play an important role when the pulse has to travel several centimeters through a preformed plasma density channel. The plasma channel is assumed to be created by a capillary discharge which generates a parabolic electron density ($n_p$) profile of the form \cite{esarey3}
\begin{equation}
n_p(r) = n_p(0) \left[1 + \left(\frac{2}{k_p r_\textnormal{ch}}\right)^2 \frac{r^2}{r_\textnormal{ch}^2}\right],
\end{equation}
where $k_p = 2 \pi/\lambda_p$ is the plasma wavenumber and $r_\textnormal{ch}$ is the channel radius. A Gaussian laser pulse of nonrelativistic intensity ($a_0^2 \ll 1$) and low peak power ($P \ll P_c$) is guided in such a plasma channel with constant radius $w_0$, if $w_0$ is matched to the channel, which means that $r_\textnormal{ch} = w_0$. Without such matching, the spot size of the pulse will oscillate during propagation. When the power of the laser pulse is of the order of the critical power, which is considered in this article, the refractive index becomes a function of intensity through the gamma-factor of the plasma electrons, which leads to self-focusing \cite{esarey3}. On the other hand, also the plasma wave generated by the laser pulse has a focusing and defocusing effect on the pulse. The front of the pulse, where the electron density has increased, experiences a weakening of the total focusing, while the back of the pulse, where the density has decreased, experiences an enhancement of the total focusing \cite{esarey3}.
In this case, self-focusing and plasma wave guiding would cause the pulse radius to oscillate even when initially matched to the channel. As a result, also the peak intensity of the pulse and the peak amplitude of the generated wakefield would change upon propagation. This would lead to a larger energy spread in the trapped electron bunch, as is shown below. Therefore it is desirable that the named oscillations are as small as possible. We found that this can be obtained by choosing the channel radius larger than required from the matching condition, $r_\textnormal{ch} = w_0$, such that self-focusing is compensated with a reduced focusing (guiding) effect of the channel.

\begin{figure}
\includegraphics[width=0.5\textwidth]{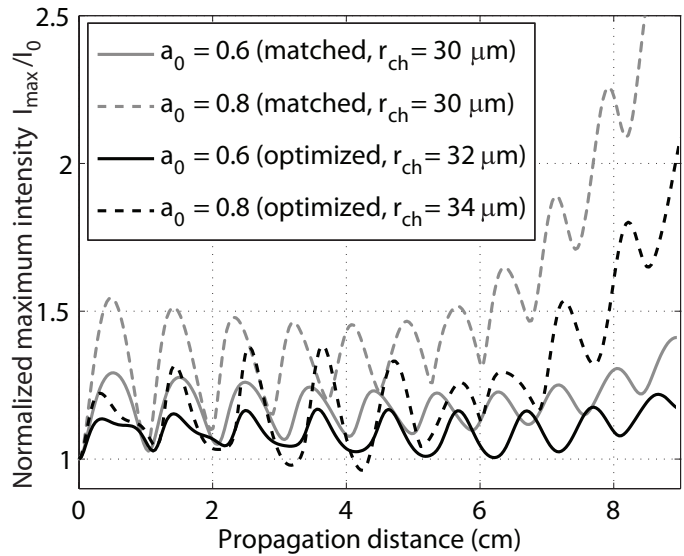}
\caption{\label{fig:normalized_amplitude}The maximum laser pulse intensity $I_\textnormal{max}$, normalized to its initial value $I_0$ as function of the propagation distance of the laser pulse through the plasma channel.}
\end{figure}

We calculated the dynamics for laser pulses with $a_0 = 0.6$ and $a_0 = 0.8$ and the formed wakefields with the fully relativistic particle code WAKE \cite{antonsen, mora}. The normalized peak intensity, as function of the propagation distance, is plotted in figure \ref{fig:normalized_amplitude}. The grey lines represent the case of a matched channel radius and the black lines represent the case of a channel with a somewhat larger (optimized) radius.
One can see that for a channel with a larger radius than for exact matching, the amplitude of the oscillations in laser intensity are essentially smaller. We found that even wider channels had increased the oscillations again. As expected, the oscillations become stronger when the peak intensity is increased and the power comes closer to the critical power.

In the region of the laser pulse, the plasma density and so also the refractive index is varying, which causes an additional temporal deformation of the pulse \cite{gordon2}. Specifically, the laser pulse is compressed and its peak intensity is increasing due to the group velocity dispersion. This pulse compression is responsible for the gradual increase of the peak intensity as can be seen in figure \ref{fig:normalized_amplitude}, and for longer propagation distances, the intensity increase is stronger. A drawback of pulse shortening on the wakefield is that it decreases the dephasing length. The laser pulse slips backwards, which causes also the wakefield to slip backwards, such that the maximum energy of an accelerated electron bunch, which can be potentially obtained, decreases.

Another effect which needs to be considered, and which was observed in the calculations, is that, during propagation of the laser pulse through the plasma, the pulse loses some of its energy, due to the generation of the wakefield. We observed that, for $a_0 = 0.6$, this loss amounts to about 4\% after 6 cm of propagation through the plasma channel. For $a_0 = 0.8$ we found a loss of about 10\% over the same propagation distance.

\section{\label{sec:injectionatangledynamics}Effect of laser pulse dynamics on injection at an angle}
\begin{figure}
\includegraphics[width=0.5\textwidth]{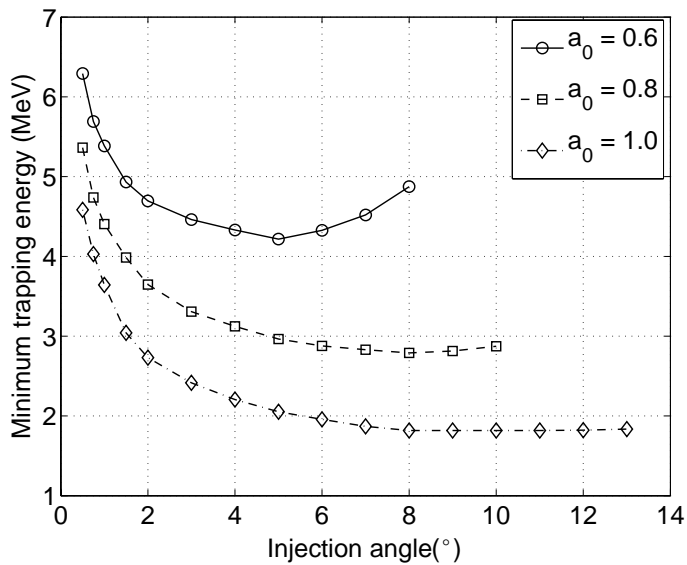}
\caption{\label{fig:angle_minimum_trapping_energy}The minimum kinetic trapping energy for the second accelerating region of the wakefield as function of the injection angle.}
\end{figure}

In this section we will investigate the dynamics of an electron bunch injected at an angle. Here we also include the full dynamics of the laser pulse. We consider wakefields generated by the laser pulses with the parameters chosen in the previous section. The dynamics of the electrons are calculated from the equation of motion with our particle tracer code. The electrons are trapped in the accelerating phases of the laser wakefield, while the focusing component of the wakefield transversely traps the electrons near the wakefield axis, where the accelerating field is at maximum. When an electron is injected into a laser wakefield, there is a minimum energy for the electron below which it cannot be trapped (see e.g. \cite{khachatryan6, khachatryan, khachatryan2}). We calculated the minimum trapping energy for the second accelerating region of the wakefield. The second accelerating region was chosen, because electrons injected at small angles can still be trapped here without going through the laser pulse which, essentially, forms a combination of injection at an angle and injection in front of the laser pulse. The same parameters as in the previous section were chosen for the laser pulse and the plasma channel. The calculation has been done for three different normalized peak amplitudes, namely $a_0 = 0.6$, $a_0 = 0.8$ and $a_0 = 1.0$, and $r_{ch} = w_0 = 30$ $\mu$m. For this set of calculations, the laser pulse dynamics was not taken into account, because the trapping time for an electron is much shorter than the typical time on which the laser pulse evolves. However, trapping of an electron bunch takes place during a time period on which laser dynamics may play an important role, as we will see below. The minimum trapping energy is plotted as function of the injection angle in figure \ref{fig:angle_minimum_trapping_energy}. One can see that the minimum trapping energy decreases when the laser pulse amplitude is increased i.e., a stronger field allows the trapping of less energetic electrons. One also sees that the minimum trapping energy decreases for larger angles. However, for $a_0 = 0.6$ the minimum trapping energy increases again when the angle becomes larger than 5 degrees and a weak increase is seen also for $a_0 = 0.8$ beyond 8 degrees. Because the injected electrons are slower than the laser pulse, they will slip backwards relative to the wakefield. If the injection angle is too small, the electrons will approach only slowly the wakefield axis while slipping backwards and this slow approach increases the chance to be scattered away from the wakefield in the defocusing region. Therefore, for a given laser wakefield and injection energy, there is a minimum injection angle for trapping. For larger injection angles the transverse momentum of the electrons is also larger and the faster approach to the axis avoids much of this scattering. However, beyond some angle the transverse momentum becomes too large for a transverse trapping and the electrons will just pass through the wakefield and cannot be trapped longitudinally and accelerated. Therefore, there is also a maximum injection angle for trapping. That an electron requires a minimum longitudinal momentum to be trapped in the wakefield can explain the increase in minimum trapping energy for a normalized amplitude of $a_0 = 0.6$ and $a_0 = 0.8$ when the angle becomes large. A larger angle for the same injection energy means that the transverse momentum increases and the longitudinal momentum decreases. To compensate for the decrease in longitudinal momentum the injection energy has to be increased, which explains why at a certain angle the minimum trapping energy increases again.

For the calculations of the dynamics of an electron bunch we use a bunch with a Gaussian density distribution in both longitudinal ($z$) and transverse ($x$ and $y$) directions. The bunch has a full-width-at-half-maximum (FWHM) duration of 250 fs (75 $\mu$m). The FWHM size of the bunch is 67 $\mu$m in both transverse directions. The parameters of the bunch are chosen in such a way that its center would move through the wakefield axis at about 120 $\mu$m behind the laser pulse. In this case electrons can be trapped simultaneously in the first, second, third and fourth accelerating regions in the wakefield. The bunch dynamics is calculated for several injection positions in the plasma channel. The longitudinal injection position of the bunch in the plasma channel is defined as the distance from the channel entrance, at which the edge of the bunch enters the wakefield. More specifically, where the bunch center is positioned at a distance $2 \sigma_b + w_0$ from the wakefield axis, where $\sigma_b$ is the rms bunch radius. As we saw in section \ref{sec:laserdynamics}, the laser pulse dynamics makes the wakefield change as function of injection position. This means that an electron bunch entering the wakefield at a different laser propagation distance in the channel (see section \ref{sec:laserdynamics}) will experience a different wakefield.

\begin{figure}
\includegraphics[width=0.5\textwidth]{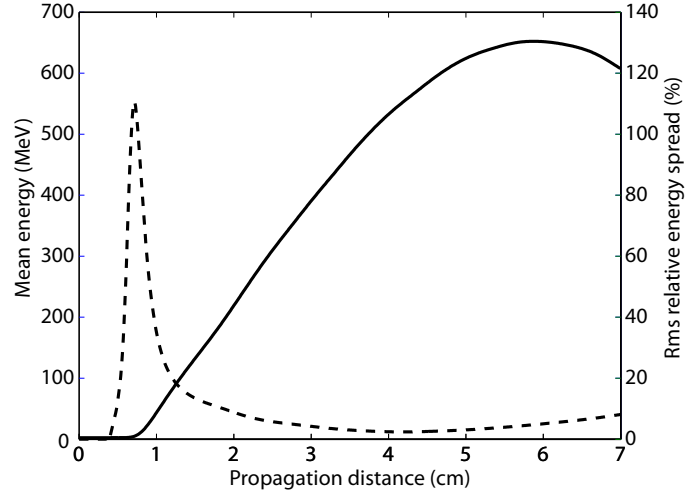}
\caption{\label{fig:energy_energy_spread}The mean energy (solid line) and rms relative energy spread (dashed line) for an electron bunch injected with a kinetic energy of 2.6 MeV, 1 cm from the entrance of the plasma channel at an angle of $\alpha=4^\circ$ into the wakefield as described in section \ref{sec:laserdynamics} with a normalized amplitude of $a_0 = 0.8$.}
\end{figure}

In the following, we present the results of our calculations of the electron dynamics in the wakefield with the parameters described in section \ref{sec:laserdynamics}. A typical example of the behaviour of the mean energy and rms relative energy spread as function of the propagation distance for an electron bunch injected at an angle is shown in figure \ref{fig:energy_energy_spread}. Here a bunch with a kinetic energy of 2.6 MeV is injected at an angle of 4 degrees at 1 cm from the entrance of the plasma channel into the wakefield described in section \ref{sec:laserdynamics} with $a_0 = 0.8$. We notice that in this case maximum energy and minimum energy spread (2.4\%) do not coincide. In the rest of this section we will vary several parameters of the problem and focus on the behaviour of the collection efficiency and minimum relative energy spread.

\begin{figure*}
\includegraphics[width=0.92\textwidth]{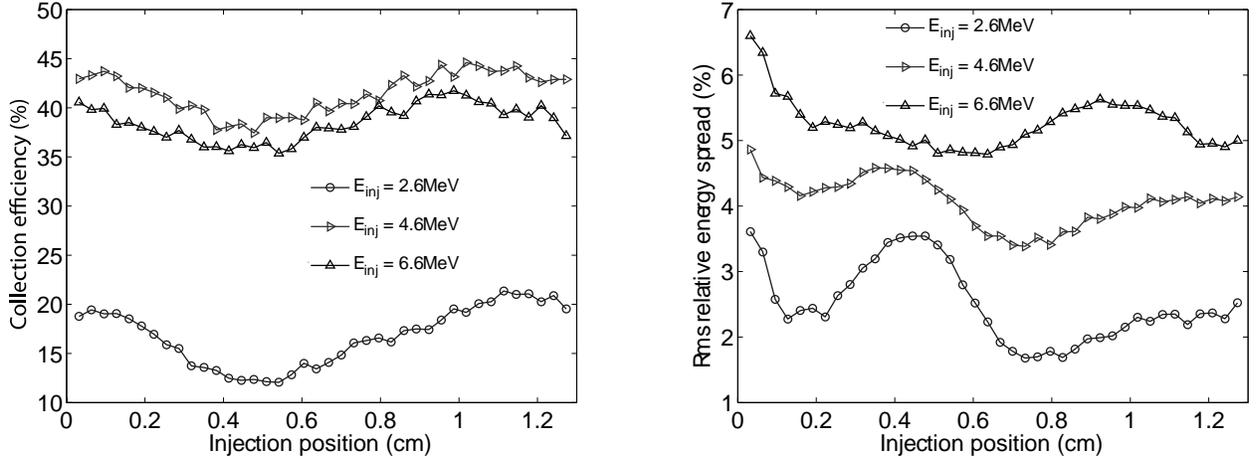}
\caption{\label{fig:a0=0_6_angle=2}The minimum rms relative energy spread and collection efficiency of the accelerated bunches versus the injection position of the externally injected electron bunch. In this case $a_0 = 0.6$ and $\alpha = 2^\circ$. The curves are shown for kinetic injection energies of 2.6, 4.6 and 6.6 MeV.}
\end{figure*}

\begin{figure*}
\includegraphics[width=0.92\textwidth]{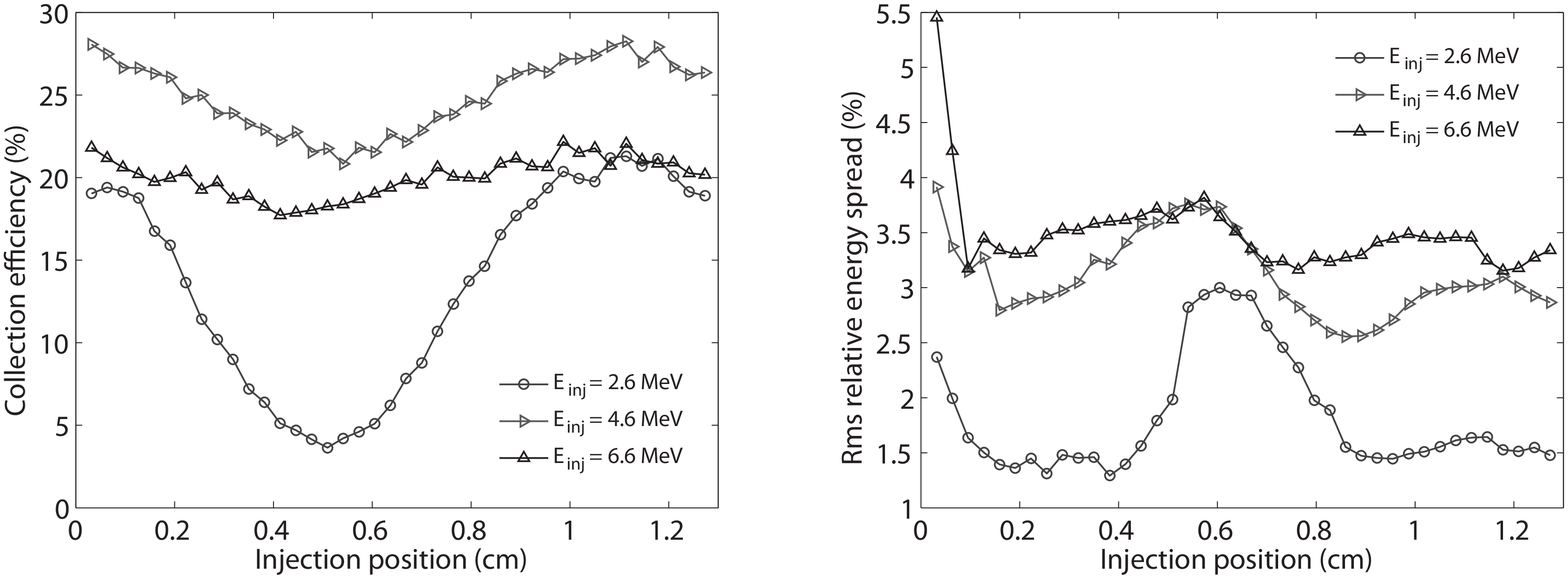}
\caption{\label{fig:a0=0_6_angle=4}The minimum rms relative energy spread and collection efficiency of the accelerated bunches versus the injection position of the externally injected electron bunch. In this case $a_0 = 0.6$ and $\alpha = 4^\circ$. The curves are shown for kinetic injection energies of 2.6, 4.6 and 6.6 MeV.}
\end{figure*}

\begin{figure*}
\includegraphics[width=0.92\textwidth]{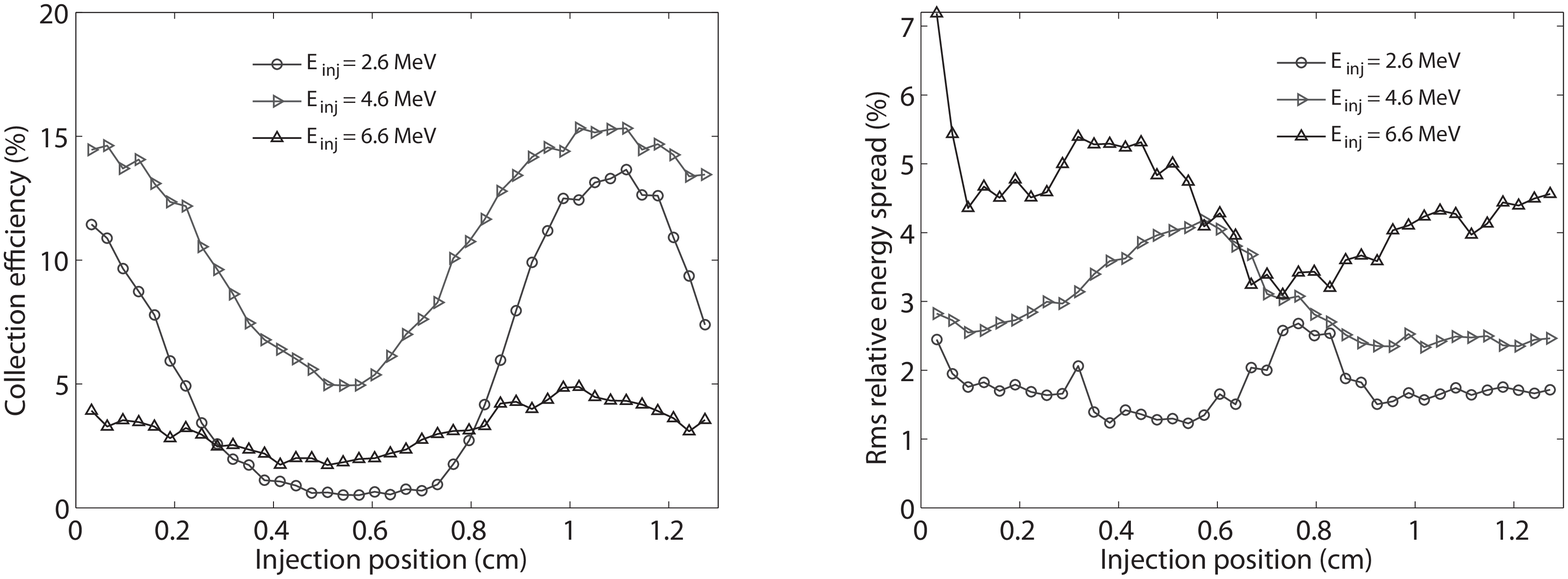}
\caption{\label{fig:a0=0_6_angle=6}The minimum rms relative energy spread and collection efficiency of the accelerated bunches versus the injection position of the externally injected electron bunch. In this case $a_0 = 0.6$ and $\alpha = 6^\circ$. The curves are shown for kinetic injection energies of 2.6, 4.6 and 6.6 MeV.}
\end{figure*}

First, we present the results obtained with the initial normalized amplitude of the laser pulse being $a_0 = 0.6$, with the plasma channel radius set to a value of 32 $\mu$m, which is a bit larger than the matched radius. The length of the plasma channel is chosen in such a way that the accelerated electron bunch has minimum relative energy spread, which turns out to occur typically between 5 and 7 cm propagation. The collection efficiency and minimum rms relative energy spread are plotted as function of the injection position (that is for different laser propagation distances, see figure \ref{fig:normalized_amplitude}) for several different injection energies and for angles of 2, 4 and 6 degrees in figure \ref{fig:a0=0_6_angle=2} to \ref{fig:a0=0_6_angle=6}.
One can see that the laser pulse dynamics plays an important role, because the energy spread and also the collection efficiency depend on the injection position. Specifically, the relative energy spread oscillates as function of the injection position. The explanation of this is based on the intensity oscillations described in section \ref{sec:laserdynamics}. For an injection position where the gradient of the laser peak intensity (see figure \ref{fig:normalized_amplitude}) is low, one obtains a low energy spread while, for injection positions with a high gradient of the peak intensity, the energy spread becomes larger. The phase of the wakefield, where an electron is trapped, depends particularly on the wakefield amplitude \cite{gordon, khachatryan6}, which is given by the peak intensity and duration of the laser pulse. If the strength of the wakefield changes while the bunch is being trapped, the electrons are trapped at different phases in the wakefield, which leads to a larger energy spread compared to the case of no laser dynamics. Correspondingly, for a low energy spread, either the laser pulse dynamics should be kept weak while the electron bunch is trapped, or the trapping time should be kept small. It can be seen that the collection efficiency also depends in some similar, oscillating manner on the position in the plasma channel where the electron bunch is injected. The explanation is that at positions where the laser intensity is higher, a stronger wakefield is generated, which can trap more electrons. 

The collection efficiency is maximum for an injection energy that is not too high, but also not too low, depending on the injection angle. For too low injection energies electrons cannot be trapped because their initial energy is below the minimum trapping energy, and for too large injection energies the transverse momentum of the electrons is too large for trapping by the transverse focusing forces. So, for an optimum collection efficiency the injection energy should not be too small, but also not too large.

The figures show that a lower energy spread is obtained for lower injection energies and that a larger injection angle can also give a lower energy spread. We observed that the trapping distance for an electron bunch, which is defined as the distance travelled by the laser pulse from the point where the first electron is trapped to the point where the last electron is trapped, decreases for larger injection angles. This explains the lower energy spread for injection with a larger angle, because the longer it takes to trap the bunch the larger the energy spread becomes \cite{khachatryan2}. On the other hand, lower injection energies and larger angles decrease the collection efficiency because there are less positions where electrons can be trapped. Thus, a compromise has to be made between collection efficiency and energy spread. For example an angle of 4 degrees, an injection energy of 2.6 MeV and an injection position of 1 cm gives a relatively low energy spread of approximately 1.6\% and a reasonably high collection efficiency of 20\%.

The rms transverse size of the accelerated bunches is lying in a range of 1.5 to 3 $\mu$m for $\alpha = 2^\circ$, 1.5 to 5 $\mu$m for $\alpha = 4^\circ$ and 1.5 to 6 $\mu$m for $\alpha = 6^\circ$. For all angles, the final energy is in the range between 300 and 500 MeV, where the final energy grows monotonically with decreasing injection energy.

A second calculation has been done with the same parameters as above, but now with the higher laser pulse intensity, namely for $a_0 = 0.8$ and a plasma channel radius of 34 $\mu$m.
\begin{figure*}
\includegraphics[width=0.92\textwidth]{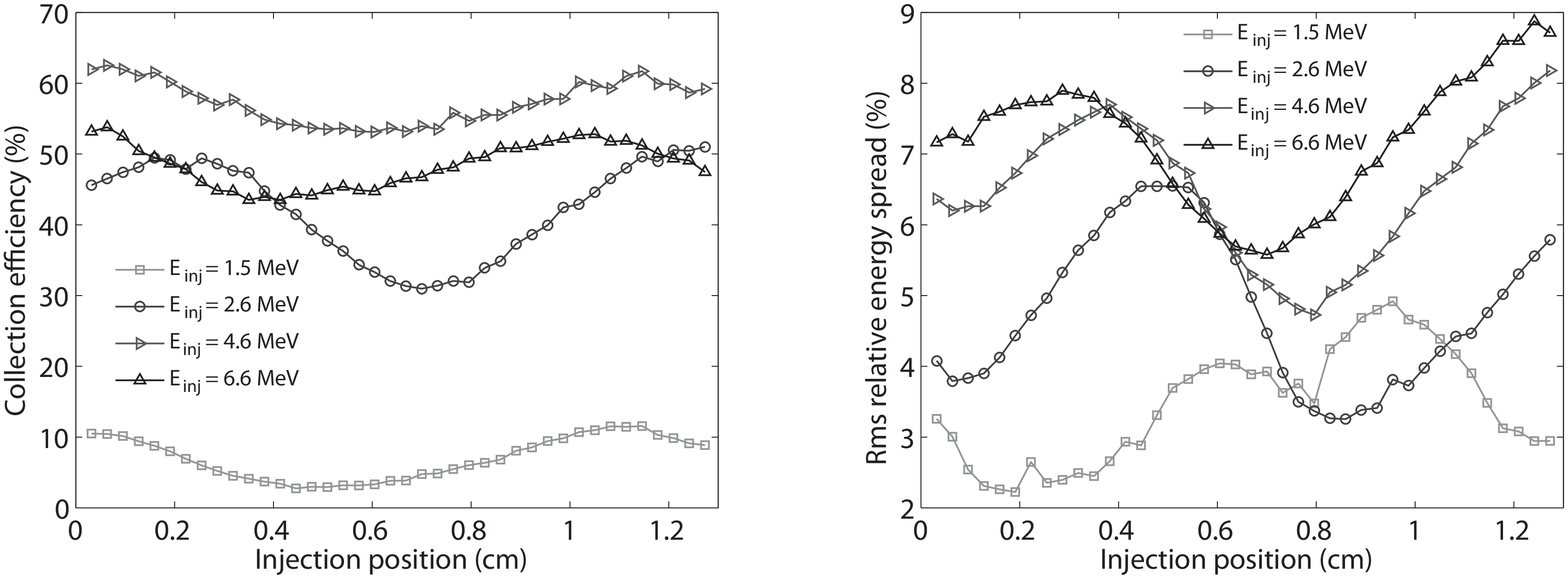}
\caption{\label{fig:a0=0_8_angle=2}The minimum rms relative energy spread and collection efficiency of the accelerated bunches versus the injection position of the externally injected electron bunch. In this case $a_0 = 0.8$ and $\alpha = 2^\circ$. The curves are shown for kinetic injection energies of 1.5, 2.6, 4.6 and 6.6 MeV.}
\end{figure*}
\begin{figure*}
\includegraphics[width=0.92\textwidth]{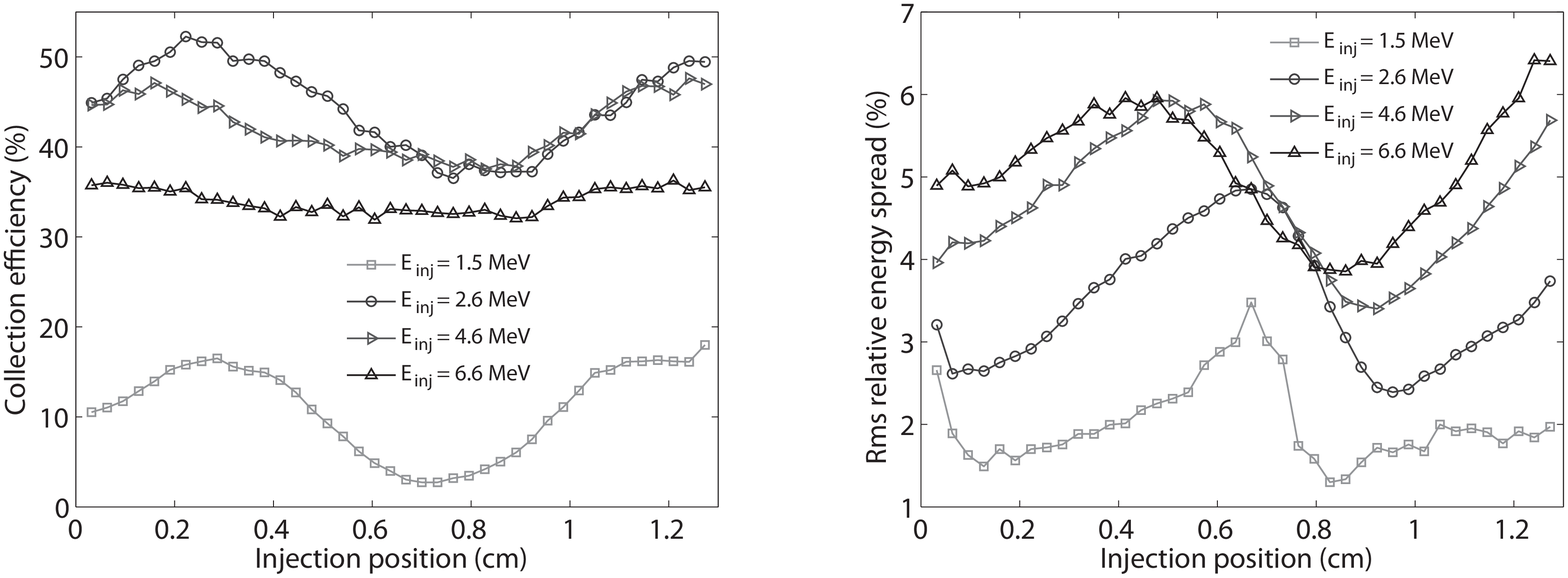}
\caption{\label{fig:a0=0_8_angle=4}The minimum rms relative energy spread and collection efficiency of the accelerated bunches versus the injection position of the externally injected electron bunch. In this case $a_0 = 0.8$ and $\alpha = 4^\circ$. The curves are shown for kinetic injection energies of 1.5, 2.6, 4.6 and 6.6 MeV.}
\end{figure*}
\begin{figure*}
\includegraphics[width=0.92\textwidth]{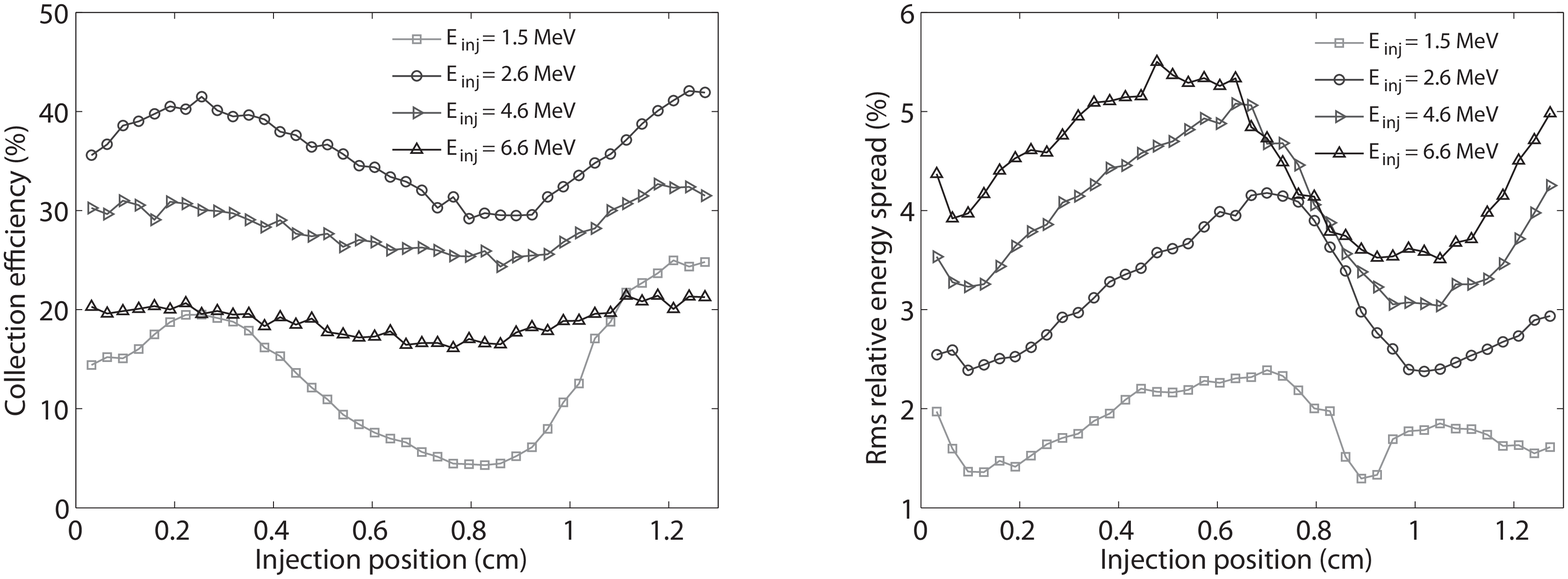}
\caption{\label{fig:a0=0_8_angle=6}The minimum rms relative energy spread and collection efficiency of the accelerated bunches versus the injection position of the externally injected electron bunch. In this case $a_0 = 0.8$ and $\alpha = 6^\circ$. The curves are shown for kinetic injection energies of 1.5, 2.6, 4.6 and 6.6 MeV.}
\end{figure*}
The results are plotted in figures \ref{fig:a0=0_8_angle=2} to \ref{fig:a0=0_8_angle=6}, where we show again the collection efficiency and energy spread for different injection energies and for injection angles of 2, 4 and 6 degrees. In general, because the wakefield in this case is stronger than in the previous case, it can be seen that more electrons can be trapped for the same injection energy. What can also be noticed is that the minimum trapping energy is lower than in the previous case (with $a_0 = 0.6$), which means that electrons even with a rather low initial kinetic energy of 1.5 MeV can be trapped.

Qualitatively, one can observe the same behavior as in the previous case, however because the dynamics of the laser pulse is stronger, some differences can be seen. The variations in the collection efficiency and the energy spread are larger, which makes it even more important to inject at an optimum position in the plasma channel. But then, e.g. by choosing the right angle, injection energy and injection position, one can obtain accelerated bunches with less than 2\% energy spread and a good collection efficiency of 20 to 30\%.

A clear difference to the previous case ($a_0 = 0.6$) is the behavior of the energy spread for an injection angle of 2 degrees (figure \ref{fig:a0=0_8_angle=2}) and an injection energy of 1.5 MeV as compared to the other angles and injection energies. We found that this is caused by a combination of injection in front of the laser pulse and injection at an angle. Electrons are trapped about 120 $\mu$m behind the laser pulse and because the injection energy is low and the angle is small, part of the bunch goes through the laser pulse in the plasma.

The energy of the accelerated bunches with $a_0 = 0.8$ is generally higher than with $a_0 = 0.6$, values between 500 and 800 MeV are obtained. Also in this case, a lower final energy corresponds to a higher injection energy. The rms transverse size of the accelerated bunches is in the range of 1 to 3 $\mu$m for $\alpha = 2^\circ$ and 1 to 3.5 $\mu$m for $\alpha = 4^\circ$ and $\alpha = 6^\circ$.

\section{\label{sec:summary} Summary and conclusion}
In this article we have studied the external injection of an electron bunch at an angle into the channel-guided laser wakefield including the dynamics of the laser pulse in the plasma channel. It turns out that the dynamics of the laser pulse has a large influence on the final energy spread of the accelerated electron bunches. Wakefields created with higher power laser pulses have stronger dynamics and the effect on the trapping and acceleration of the electrons is also stronger. However by varying the injection position, injection energy and injection angle, the energy spread can be minimized, while higher collection efficiencies can be achieved. This shows that, even when the laser pulse dynamics becomes important, micron-sized electron bunches accelerated to several hundreds of MeV's can be generated by injection at an angle. In particular minimizing the energy spread can be achieved by injecting the electron bunch at a position in the channel, where the wakefield is as constant as possible. Minimizing the trapping distance by increasing the injection angle and decreasing the injection energy can lower the energy spread even more.

The energy spread was calculated for all formed and accelerated bunches. However the energy spread for one single bunch can be considerably lower. The energy and energy spread of a bunch depends on how far behind the laser pulse it is formed. The farther from the pulse, the weaker the accelerating field and the stronger the focussing field becomes. The wakefield also gets curved \cite{esarey, khachatryan4}. The bunches that are accelerated at a greater distance from the pulse in general have a lower energy and a higher energy spread, which means that the total energy spread will increase when more bunches are formed. The injected bunch should be kept short to minimize the number of formed bunches.

We can conclude that external injection at an angle into the laser wakefield is an important alternative when the scattering of the injected bunch by a laser pulse or by the wakefield in the vacuum-plasma transition should be avoided.

\begin{acknowledgments}
We acknowledge the support of the Dutch Ministry of Education, Culture and Science (OC\&W), the Dutch Foundation for Fundamental Research on Matter (FOM) under the ''Laser wakefield accelerators'' programme and the European Community-New and Emerging Science and Technology Activity under the FP6 “Structuring the European Research Area” programme (project EuroLEAP, contract number 028514).
\end{acknowledgments}


\newpage

\begin{thebibliography}{27}
\expandafter\ifx\csname natexlab\endcsname\relax\def\natexlab#1{#1}\fi
\expandafter\ifx\csname bibnamefont\endcsname\relax
  \def\bibnamefont#1{#1}\fi
\expandafter\ifx\csname bibfnamefont\endcsname\relax
  \def\bibfnamefont#1{#1}\fi
\expandafter\ifx\csname citenamefont\endcsname\relax
  \def\citenamefont#1{#1}\fi
\expandafter\ifx\csname url\endcsname\relax
  \def\url#1{\texttt{#1}}\fi
\expandafter\ifx\csname urlprefix\endcsname\relax\def\urlprefix{URL }\fi
\providecommand{\bibinfo}[2]{#2}
\providecommand{\eprint}[2][]{\url{#2}}

\bibitem[{\citenamefont{Antonsen and Mora}(1992)}]{antonsen}
\bibinfo{author}{\bibfnamefont{T.~M.} \bibnamefont{Antonsen}}, and
  \bibinfo{author}{\bibfnamefont{P.}~\bibnamefont{Mora}}, \bibinfo{year}{1992},
  \bibinfo{journal}{Phys. Rev. Lett.}
  \textbf{\bibinfo{volume}{69}}(\bibinfo{number}{15}), \bibinfo{pages}{2204}.

\bibitem[{\citenamefont{Esarey} \emph{et~al.}(1997)\citenamefont{Esarey,
  Sprangle, J., and Ting}}]{esarey3}
\bibinfo{author}{\bibfnamefont{E.} \bibnamefont{Esarey}},
  \bibinfo{author}{\bibfnamefont{P.}~\bibnamefont{Sprangle}},
  \bibinfo{author}{\bibfnamefont{J.}~\bibnamefont{Krall}}, and
  \bibinfo{author}{\bibfnamefont{A.}~\bibnamefont{Ting}}, \bibinfo{year}{1997},
  \bibinfo{journal}{IEEE. J. Quantum Electron.} \textbf{\bibinfo{volume}{33}},
  \bibinfo{pages}{1879}.

\bibitem[{\citenamefont{Esarey} \emph{et~al.}(1996)\citenamefont{Esarey,
  Sprangle, Krall, and Ting}}]{esarey}
\bibinfo{author}{\bibfnamefont{E.} \bibnamefont{Esarey}},
  \bibinfo{author}{\bibfnamefont{P.}~\bibnamefont{Sprangle}},
  \bibinfo{author}{\bibfnamefont{J.}~\bibnamefont{Krall}}, and
  \bibinfo{author}{\bibfnamefont{A.}~\bibnamefont{Ting}}, \bibinfo{year}{1996},
  \bibinfo{journal}{IEEE Trans. Plasma Sci.}
  \textbf{\bibinfo{volume}{24}}(\bibinfo{number}{2}), \bibinfo{pages}{252}.

\bibitem[{\citenamefont{Esirkepov} \emph{et~al.}(2006)\citenamefont{Esirkepov,
  Bulanov, Yamagiwa, and Tajima}}]{esirkepov}
\bibinfo{author}{\bibfnamefont{T.} \bibnamefont{Esirkepov}},
  \bibinfo{author}{\bibfnamefont{S.~V.} \bibnamefont{Bulanov}},
  \bibinfo{author}{\bibfnamefont{M.}~\bibnamefont{Yamagiwa}}, and
  \bibinfo{author}{\bibfnamefont{T.}~\bibnamefont{Tajima}},
  \bibinfo{year}{2006}, \bibinfo{journal}{Phys. Rev. Lett.}
  \textbf{\bibinfo{volume}{96}}, \bibinfo{pages}{014803}.

\bibitem[{\citenamefont{Faure} \emph{et~al.}(2004)\citenamefont{Faure, Glinec,
  Pukhov, Kiselev, Gordienko, Lefebvre, Rousseau, Burgy, and Malka}}]{faure}
\bibinfo{author}{\bibfnamefont{J.} \bibnamefont{Faure}},
  \bibinfo{author}{\bibfnamefont{Y.}~\bibnamefont{Glinec}},
  \bibinfo{author}{\bibfnamefont{A.}~\bibnamefont{Pukhov}},
  \bibinfo{author}{\bibfnamefont{S.}~\bibnamefont{Kiselev}},
  \bibinfo{author}{\bibfnamefont{S.}~\bibnamefont{Gordienko}},
  \bibinfo{author}{\bibfnamefont{E.}~\bibnamefont{Lefebvre}},
  \bibinfo{author}{\bibfnamefont{J.~P.} \bibnamefont{Rousseau}},
  \bibinfo{author}{\bibfnamefont{F.}~\bibnamefont{Burgy}}, and
  \bibinfo{author}{\bibfnamefont{V.}~\bibnamefont{Malka}},
  \bibinfo{year}{2004}, \bibinfo{journal}{Nature}
  \textbf{\bibinfo{volume}{431}}.

\bibitem[{\citenamefont{Faure} \emph{et~al.}(2006)\citenamefont{Faure,
  Rechatin, Norlin, Lifschitz, Glinec, and Malka}}]{faure2}
\bibinfo{author}{\bibfnamefont{J.} \bibnamefont{Faure}},
  \bibinfo{author}{\bibfnamefont{C.}~\bibnamefont{Rechatin}},
  \bibinfo{author}{\bibfnamefont{A.}~\bibnamefont{Norlin}},
  \bibinfo{author}{\bibfnamefont{A.}~\bibnamefont{Lifschitz}},
  \bibinfo{author}{\bibfnamefont{Y.}~\bibnamefont{Glinec}}, and
  \bibinfo{author}{\bibfnamefont{V.}~\bibnamefont{Malka}},
  \bibinfo{year}{2006}, \bibinfo{journal}{Nature}
  \textbf{\bibinfo{volume}{444}}.

\bibitem[{\citenamefont{Geddes} \emph{et~al.}(2004)\citenamefont{Geddes, Toth,
  van Tilborg, Esarey, Schroeder, Bruhwiler, Nieter, Cary, and
  Leemans}}]{geddes}
\bibinfo{author}{\bibfnamefont{C.~G.~R.} \bibnamefont{Geddes}},
  \bibinfo{author}{\bibfnamefont{C.}~\bibnamefont{Toth}},
  \bibinfo{author}{\bibfnamefont{J.}~\bibnamefont{van Tilborg}},
  \bibinfo{author}{\bibfnamefont{E.}~\bibnamefont{Esarey}},
  \bibinfo{author}{\bibfnamefont{C.~B.} \bibnamefont{Schroeder}},
  \bibinfo{author}{\bibfnamefont{D.}~\bibnamefont{Bruhwiler}},
  \bibinfo{author}{\bibfnamefont{C.}~\bibnamefont{Nieter}},
  \bibinfo{author}{\bibfnamefont{J.}~\bibnamefont{Cary}}, and
  \bibinfo{author}{\bibfnamefont{W.~P.} \bibnamefont{Leemans}},
  \bibinfo{year}{2004}, \bibinfo{journal}{Nature}
  \textbf{\bibinfo{volume}{431}}.

\bibitem[{\citenamefont{Gorbunov} \emph{et~al.}(2005)\citenamefont{Gorbunov,
  Kalmykov, and Mora}}]{gorbunov}
\bibinfo{author}{\bibfnamefont{L.~M.} \bibnamefont{Gorbunov}},
  \bibinfo{author}{\bibfnamefont{S.~Y.} \bibnamefont{Kalmykov}}, and
  \bibinfo{author}{\bibfnamefont{P.}~\bibnamefont{Mora}}, \bibinfo{year}{2005},
  \bibinfo{journal}{Phys. Plasmas} \textbf{\bibinfo{volume}{12}}.

\bibitem[{\citenamefont{Gordon} \emph{et~al.}(2003)\citenamefont{Gordon,
  Hafizi, Hubbard, Penano, Sprangle, and Ting}}]{gordon2}
\bibinfo{author}{\bibfnamefont{D.~F.} \bibnamefont{Gordon}},
  \bibinfo{author}{\bibfnamefont{B.}~\bibnamefont{Hafizi}},
  \bibinfo{author}{\bibfnamefont{R.~F.} \bibnamefont{Hubbard}},
  \bibinfo{author}{\bibfnamefont{J.~R.} \bibnamefont{Penano}},
  \bibinfo{author}{\bibfnamefont{P.}~\bibnamefont{Sprangle}}, and
  \bibinfo{author}{\bibfnamefont{A.}~\bibnamefont{Ting}}, \bibinfo{year}{2003},
  \bibinfo{journal}{Phys. Rev. Lett.}
  \textbf{\bibinfo{volume}{90}}(\bibinfo{number}{21}).

\bibitem[{\citenamefont{Gordon} \emph{et~al.}(2005)\citenamefont{Gordon,
  Hubbard, Hafizi, Ting, and P.}}]{gordon}
\bibinfo{author}{\bibfnamefont{D.~F.} \bibnamefont{Gordon}},
  \bibinfo{author}{\bibfnamefont{J.~H.} \bibnamefont{Hubbard}},
  \bibinfo{author}{\bibfnamefont{B.}~\bibnamefont{Hafizi}},
  \bibinfo{author}{\bibfnamefont{A.}~\bibnamefont{Ting}}, and
  \bibinfo{author}{\bibfnamefont{P.}~\bibnamefont{Sprangle}}, \bibinfo{year}{2005},
  \bibinfo{journal}{Phys. Rev. E}
  \textbf{\bibinfo{volume}{71}}(\bibinfo{number}{026404}).

\bibitem[{\citenamefont{Hubbard} \emph{et~al.}(2005)\citenamefont{Hubbard,
  Gordon, Cooley, Hafizi, Jones, Kaganovich, Sprangle, Ting, and
  Zigler}}]{hubbard}
\bibinfo{author}{\bibfnamefont{R.~F.} \bibnamefont{Hubbard}},
  \bibinfo{author}{\bibfnamefont{D.~F.} \bibnamefont{Gordon}},
  \bibinfo{author}{\bibfnamefont{J.~H.} \bibnamefont{Cooley}},
  \bibinfo{author}{\bibfnamefont{B.}~\bibnamefont{Hafizi}},
  \bibinfo{author}{\bibfnamefont{T.~G.} \bibnamefont{Jones}},
  \bibinfo{author}{\bibfnamefont{D.}~\bibnamefont{Kaganovich}},
  \bibinfo{author}{\bibfnamefont{P.}~\bibnamefont{Sprangle}},
  \bibinfo{author}{\bibfnamefont{A.}~\bibnamefont{Ting}}, and
  \bibinfo{author}{\bibfnamefont{A.}~\bibnamefont{Zigler}},
  \bibinfo{year}{2005}, \bibinfo{journal}{IEEE Trans. Plasma Sci.}
  \textbf{\bibinfo{volume}{33}}(\bibinfo{number}{2}), \bibinfo{pages}{712}.

\bibitem[{\citenamefont{Irman} \emph{et~al.}(2007)\citenamefont{Irman,
  Luttikhof, Khachatryan, van Goor, Verschuur, Bastiaens, and Boller}}]{irman}
\bibinfo{author}{\bibfnamefont{A.} \bibnamefont{Irman}},
  \bibinfo{author}{\bibfnamefont{M.~J.~H.} \bibnamefont{Luttikhof}},
  \bibinfo{author}{\bibfnamefont{A.~G.} \bibnamefont{Khachatryan}},
  \bibinfo{author}{\bibfnamefont{F.~A.} \bibnamefont{van Goor}},
  \bibinfo{author}{\bibfnamefont{J.~W.~J.} \bibnamefont{Verschuur}},
  \bibinfo{author}{\bibfnamefont{H.~M.~J.} \bibnamefont{Bastiaens}}, and
  \bibinfo{author}{\bibfnamefont{K.-J.} \bibnamefont{Boller}},
  \bibinfo{year}{2007}, \bibinfo{journal}{J. Appl. Phys.}
  \textbf{\bibinfo{volume}{102}}(\bibinfo{number}{2}).

\bibitem[{\citenamefont{Kalmykov} \emph{et~al.}(2006)\citenamefont{Kalmykov,
  Grobunov, Mora, and Shvets}}]{kalmykov}
\bibinfo{author}{\bibfnamefont{S.~Y.} \bibnamefont{Kalmykov}},
  \bibinfo{author}{\bibfnamefont{L.~M.} \bibnamefont{Grobunov}},
  \bibinfo{author}{\bibfnamefont{P.}~\bibnamefont{Mora}}, and
  \bibinfo{author}{\bibfnamefont{G.}~\bibnamefont{Shvets}},
  \bibinfo{year}{2006}, \bibinfo{journal}{Phys. Plasmas}
  \textbf{\bibinfo{volume}{13}}(\bibinfo{number}{113102}).

\bibitem[{\citenamefont{Khachatryan}(1999)}]{khachatryan4}
\bibinfo{author}{\bibfnamefont{A.~G.} \bibnamefont{Khachatryan}},
  \bibinfo{year}{1999}, \bibinfo{journal}{Phys. Rev. E}
  \textbf{\bibinfo{volume}{60}}(\bibinfo{number}{5}), \bibinfo{pages}{6210}.

\bibitem[{\citenamefont{Khachatryan}(2001)}]{khachatryan5}
\bibinfo{author}{\bibfnamefont{A.~G.} \bibnamefont{Khachatryan}},
  \bibinfo{year}{2001}, \bibinfo{journal}{JETP Letters}
  \textbf{\bibinfo{volume}{74}}(\bibinfo{number}{7}), \bibinfo{pages}{371}.

\bibitem[{\citenamefont{Khachatryan}(2002)}]{khachatryan}
\bibinfo{author}{\bibfnamefont{A.~G.} \bibnamefont{Khachatryan}},
  \bibinfo{year}{2002}, \bibinfo{journal}{Phys. Rev. E}
  \textbf{\bibinfo{volume}{65}}(\bibinfo{number}{046504}).

\bibitem[{\citenamefont{Khachatryan}
  \emph{et~al.}(2004)\citenamefont{Khachatryan, van Goor, Boller, Reitsma, and
  Jaroszynski}}]{khachatryan2}
\bibinfo{author}{\bibfnamefont{A.~G.} \bibnamefont{Khachatryan}},
  \bibinfo{author}{\bibfnamefont{F.~A.} \bibnamefont{van Goor}},
  \bibinfo{author}{\bibfnamefont{K.-J.} \bibnamefont{Boller}},
  \bibinfo{author}{\bibfnamefont{A.~J.~W.} \bibnamefont{Reitsma}}, and
  \bibinfo{author}{\bibfnamefont{D.~A.} \bibnamefont{Jaroszynski}},
  \bibinfo{year}{2004}, \bibinfo{journal}{Phys. Rev. Special Topics}
  \textbf{\bibinfo{volume}{7}}(\bibinfo{number}{121301}).

\bibitem[{\citenamefont{Khachatryan}
  \emph{et~al.}(2007)\citenamefont{Khachatryan, Luttikhof, van Goor, and
  Boller}}]{khachatryan6}
\bibinfo{author}{\bibfnamefont{A.~G.} \bibnamefont{Khachatryan}},
  \bibinfo{author}{\bibfnamefont{M.~J.~H.} \bibnamefont{Luttikhof}},
  \bibinfo{author}{\bibfnamefont{F.~A.} \bibnamefont{van Goor}}, and
  \bibinfo{author}{\bibfnamefont{K.-J.} \bibnamefont{Boller}},
  \bibinfo{year}{2007}, \bibinfo{journal}{Appl. Phys. B}
  \textbf{\bibinfo{volume}{86}}(\bibinfo{number}{1}).

\bibitem[{\citenamefont{Khachatryan}
  \emph{et~al.}(2006)\citenamefont{Khachatryan, Luttikhof, Irman, van Goor,
  Verschuur, Bastiaens, and Boller}}]{khachatryan7}
\bibinfo{author}{\bibfnamefont{A.~G.} \bibnamefont{Khachatryan}},
  \bibinfo{author}{\bibfnamefont{M.~J.~H.} \bibnamefont{Luttikhof}},
  \bibinfo{author}{\bibfnamefont{A.}~\bibnamefont{Irman}},
  \bibinfo{author}{\bibfnamefont{F.~A.} \bibnamefont{van Goor}},
  \bibinfo{author}{\bibfnamefont{J.~W.~J.} \bibnamefont{Verschuur}},
  \bibinfo{author}{\bibfnamefont{H.~M.~J.} \bibnamefont{Bastiaens}}, and
  \bibinfo{author}{\bibfnamefont{K.-J.} \bibnamefont{Boller}},
  \bibinfo{year}{2006}, \bibinfo{journal}{Nucl. Instrum. Methods A}
  \textbf{\bibinfo{volume}{566}}, \bibinfo{pages}{244}.

\bibitem[{\citenamefont{Leemans} \emph{et~al.}(2006)\citenamefont{Leemans,
  Nagler, Gonsalves, T\'{o}th, Nakamura, Geddes, Esarey, Schroeder, and
  Hooker}}]{leemans}
\bibinfo{author}{\bibfnamefont{W.~P.} \bibnamefont{Leemans}},
  \bibinfo{author}{\bibfnamefont{B.}~\bibnamefont{Nagler}},
  \bibinfo{author}{\bibfnamefont{A.~J.} \bibnamefont{Gonsalves}},
  \bibinfo{author}{\bibfnamefont{C.}~\bibnamefont{T\'{o}th}},
  \bibinfo{author}{\bibfnamefont{K.}~\bibnamefont{Nakamura}},
  \bibinfo{author}{\bibfnamefont{C.~G.~R.} \bibnamefont{Geddes}},
  \bibinfo{author}{\bibfnamefont{E.}~\bibnamefont{Esarey}},
  \bibinfo{author}{\bibfnamefont{C.~B.} \bibnamefont{Schroeder}}, and
  \bibinfo{author}{\bibfnamefont{S.~M.} \bibnamefont{Hooker}},
  \bibinfo{year}{2006}, \bibinfo{journal}{Nature Phys.}
  \textbf{\bibinfo{volume}{2}}, \bibinfo{pages}{696}.

\bibitem[{\citenamefont{Lifschitz} \emph{et~al.}(2006)\citenamefont{Lifschitz,
  Faure, Glinec, Malka, and Mora}}]{lifschitz2}
\bibinfo{author}{\bibfnamefont{A.~F.} \bibnamefont{Lifschitz}},
  \bibinfo{author}{\bibfnamefont{J.}~\bibnamefont{Faure}},
  \bibinfo{author}{\bibfnamefont{Y.}~\bibnamefont{Glinec}},
  \bibinfo{author}{\bibfnamefont{V.}~\bibnamefont{Malka}}, and
  \bibinfo{author}{\bibfnamefont{P.}~\bibnamefont{Mora}}, \bibinfo{year}{2006},
  \bibinfo{journal}{Laser Part. Beams} \textbf{\bibinfo{volume}{24}},
  \bibinfo{pages}{255}.

\bibitem[{\citenamefont{Lifschitz} \emph{et~al.}(2005)\citenamefont{Lifschitz,
  Faure, Malka, and Mora}}]{lifschitz}
\bibinfo{author}{\bibfnamefont{A.~F.} \bibnamefont{Lifschitz}},
  \bibinfo{author}{\bibfnamefont{J.}~\bibnamefont{Faure}},
  \bibinfo{author}{\bibfnamefont{V.}~\bibnamefont{Malka}}, and
  \bibinfo{author}{\bibfnamefont{P.}~\bibnamefont{Mora}}, \bibinfo{year}{2005},
  \bibinfo{journal}{Phys. Plasmas} \textbf{\bibinfo{volume}{12}}.

\bibitem[{\citenamefont{Luttikhof} \emph{et~al.}(2007)\citenamefont{Luttikhof,
  Khachatryan, van Goor, and Boller}}]{luttikhof}
\bibinfo{author}{\bibfnamefont{M.~J.~H.} \bibnamefont{Luttikhof}},
  \bibinfo{author}{\bibfnamefont{A.~G.} \bibnamefont{Khachatryan}},
  \bibinfo{author}{\bibfnamefont{F.~A.} \bibnamefont{van Goor}}, and
  \bibinfo{author}{\bibfnamefont{K.-J.} \bibnamefont{Boller}},
  \bibinfo{year}{2007}, \bibinfo{journal}{Phys. Plasmas}
  \textbf{\bibinfo{volume}{14}}(\bibinfo{number}{8}).

\bibitem[{\citenamefont{Mangles} \emph{et~al.}(2004)\citenamefont{Mangles,
  Murphy, Najmudin, Thomas, Collier, Dangor, Divall, Foster, Gallacher, Hooker,
  Jaroszynski, Langley} \emph{et~al.}}]{mangles}
\bibinfo{author}{\bibfnamefont{S.~P.~D.} \bibnamefont{Mangles}},
  \bibinfo{author}{\bibfnamefont{C.~D.} \bibnamefont{Murphy}},
  \bibinfo{author}{\bibfnamefont{Z.}~\bibnamefont{Najmudin}},
  \bibinfo{author}{\bibfnamefont{A.~G.~R.} \bibnamefont{Thomas}},
  \bibinfo{author}{\bibfnamefont{J.~L.} \bibnamefont{Collier}},
  \bibinfo{author}{\bibfnamefont{A.~E.} \bibnamefont{Dangor}},
  \bibinfo{author}{\bibfnamefont{E.~J.} \bibnamefont{Divall}},
  \bibinfo{author}{\bibfnamefont{P.~S.} \bibnamefont{Foster}},
  \bibinfo{author}{\bibfnamefont{J.~G.} \bibnamefont{Gallacher}},
  \bibinfo{author}{\bibfnamefont{C.~J.} \bibnamefont{Hooker}},
  \bibinfo{author}{\bibfnamefont{D.~A.} \bibnamefont{Jaroszynski}},
  \bibinfo{author}{\bibfnamefont{A.~J.} \bibnamefont{Langley}}, \emph{et~al.},
  \bibinfo{year}{2004}, \bibinfo{journal}{Nature}
  \textbf{\bibinfo{volume}{431}}.

\bibitem[{\citenamefont{Mora and Antonsen}(1996)}]{mora}
\bibinfo{author}{\bibfnamefont{P.} \bibnamefont{Mora}}, and
  \bibinfo{author}{\bibfnamefont{T.}~\bibnamefont{Antonsen}},
  \bibinfo{year}{1996}, \bibinfo{journal}{Phys. Rev. E.}
  \textbf{\bibinfo{volume}{53}}(\bibinfo{number}{3}), \bibinfo{pages}{R2068}.

\bibitem[{\citenamefont{Pukhov and Meyer-ter Vehn}(2002)}]{pukhov}
\bibinfo{author}{\bibfnamefont{A.} \bibnamefont{Pukhov}}, and
  \bibinfo{author}{\bibfnamefont{J.}~\bibnamefont{Meyer-ter Vehn}},
  \bibinfo{year}{2002}, \bibinfo{journal}{Appl. Phys. B}
  \textbf{\bibinfo{volume}{74}}(\bibinfo{number}{4-5}), \bibinfo{pages}{355}.

\bibitem[{\citenamefont{Tajima and Dawson}(1979)}]{tajima}
\bibinfo{author}{\bibfnamefont{T.} \bibnamefont{Tajima}}, and
  \bibinfo{author}{\bibfnamefont{J.~M.} \bibnamefont{Dawson}},
  \bibinfo{year}{1979}, \bibinfo{journal}{Phys. Rev. Lett.}
  \textbf{\bibinfo{volume}{43}}(\bibinfo{number}{4}), \bibinfo{pages}{267}.

\bibitem[{\citenamefont{Urbanus} \emph{et~al.}(2006)\citenamefont{Urbanus, van
  Dijk, van~der Geer, Brussaard, and van~der Wiel}}]{urbanus}
\bibinfo{author}{\bibfnamefont{W.~H.} \bibnamefont{Urbanus}},
  \bibinfo{author}{\bibfnamefont{W.}~\bibnamefont{van Dijk}},
  \bibinfo{author}{\bibfnamefont{S.~B.} \bibnamefont{van~der Geer}},
  \bibinfo{author}{\bibfnamefont{G.~J.~H.} \bibnamefont{Brussaard}}, and
  \bibinfo{author}{\bibfnamefont{M.~J.} \bibnamefont{van~der Wiel}},
  \bibinfo{year}{2006}, \bibinfo{journal}{J. Appl. Phys.}
  \textbf{\bibinfo{volume}{99}}(\bibinfo{number}{11}).

\end{thebibliography}

\end{document}